# Language of physics, language of math:
# Disciplinary culture and dynamic epistemology

Edward F. Redish*
Department of Physics,
University of Maryland, College Park, MD 20742

Eric Kuo
Department of Physics & Graduate School of Education
Stanford University, Stanford, CA 94305

## Abstract:

Mathematics is a critical part of much scientific research. Physics in particular weaves math extensively into its instruction beginning in high school. Despite much research on the learning of both physics and math, the problem of how to effectively include math in physics in a way that reaches most students remains unsolved. In this paper, we suggest that a fundamental issue has received insufficient exploration: the fact that in science, we don't just use math, we make meaning with it in a different way than mathematicians do. In this reflective essay, we explore math as a language and consider the language of math in physics through the lens of cognitive linguistics. We begin by offering a number of examples that show how the use of math in physics differs from the use of math as typically found in math classes. We then explore basic concepts in cognitive semantics to show how humans make meaning with language in general. The critical elements are the roles of embodied cognition and interpretation in context. Then we show how a theoretical framework commonly used in physics education research, resources, is coherent with and extends the ideas of cognitive semantics by connecting embodiment to phenomenological primitives and contextual interpretation to the dynamics of meaning making with conceptual resources, epistemological resources, and affect. We present these ideas with illustrative case studies of students working on physics problems with math and demonstrate the dynamical nature of student reasoning with math in physics. We conclude with some thoughts about the implications for instruction.

* Corresponding author:
Email: redish@umd.edu
Phone: +1-301-405-6120
Fax: +1-301-314-9531

# 1    Introduction / Problem statement

Mathematics is deeply woven into both the teaching and practice of physics.[1]  Other sciences, such as chemistry, biology, geology, and meteorology often use math extensively in practice, but typically rely less heavily on it than physics does in high school and college instruction. Despite decades of experience in teaching physics with math from the high school to the graduate level, and despite years of research on "problem solving" in physics (Maloney 1994, Hsu et al. 2004), the physics community continues to have mixed success in teaching students to use math effectively in physics. A few students seem to take naturally and comfortably to the use of math to describe the physical world; but many struggle with it, both at the introductory and at the more advanced level, even though multiple math classes may be required as pre-requisites for physics classes. For example, a physics major at the University of Maryland is required to take five semesters of mathematics and those intending to go on to graduate school are encouraged to take at least two more.

One of us (EFR) has been a faculty member in a large research university physics department for more than forty years. In this time he has observed or participated in four major re-considerations of the physics major's curriculum. In each of these, a critical element of the discussion has been the question, "Why do so many students seem unable to use math in our physics classes, despite having had success in their pre-requisite math classes?" This question is raised at all levels, from the introductory ("They can't do simple algebra!"), to the upper division ("They have terrible trouble with simple matrix concepts!"), to the graduate ("Many of them can't take a simple Fourier transform!").

In this reflective essay, we consider the issue of the use of mathematics as a language in science.

First, we argue that there are dramatic, if often unrecognized, differences between the disciplinary cultures of mathematics and physics in how they use and interpret mathematical expressions. In section 2, we demonstrate through examples how the interpretation of math by physics instructors can blend in physical knowledge, changing the interpretation of the math. The result is that the equations that students learn in their math classes may look and feel very different from similar equations that they see in their science classes (especially physics). Some of these differences are deep. In section 3, we reach out to the disciplines of linguistics and semantics to explore the fundamental processes associated with meaning-making in the use of everyday language and bridge these processes of meaning-making to the use of mathematics.

The key point is this: Although the formal mathematical syntax may be the same across the disciplines of mathematics and physics, the uses and meanings of that formal syntax may differ dramatically between the two disciplines. These differences in semantic meaning may be masked by the apparent similarity in the formal syntax.

Second, we seek to understand how students manage these multiple languages across the disciplines. In section 4, we turn to making meaning with math in physics by briefly summarizing the resources theoretical framework of student cognition. This is based on a Knowledge-in-Pieces model modulated by dynamic framing through student expectations, epistemology, and affect. We consider in detail a few case studies of students reasoning with equations in physics, showing how knowledge, epistemology, and affect interact dynamically to make different kinds of meaning for students. Section 5 gives a brief discussion of the implications for instruction.

## 2    Math is different in a physics context

Math-in-science (and particularly math-in-physics) is not the same as doing math. It has a different purpose – representing meaning about physical systems rather than expressing abstract relationships. It even has a distinct semiotics – the way meaning is put into symbols – from pure mathematics.

---

[1]  Parts of this paper are based on contributions to conference proceedings (Redish 2005, Redish & Gupta 2009).

It almost seems that the "language" of mathematics we use in physics is not the same as the one taught by mathematicians. There are many important differences in what seems to be the physicist's "dialect" of speaking math, so, while related, the languages of "math in math" and "math in physics" may need to be considered as separate languages. The key difference is that loading physical meaning onto symbols does work for physicists and leads to differences in how physicists and mathematicians interpret equations. We not only *use math in doing physics*, we *use physics in doing math*. We present three examples illustrating different aspects of the cultural differences between the use of math by physicists and mathematicians and then discuss the general structure of mapping meaning to math.

## 2.1 Loading meaning onto symbols leads to differences in how physicists and mathematicians interpret equations

### 2.1.1 Corinne's Shibboleth

Our first example is "Corinne's Shibboleth".[2] (Dray & Manogoue, 2002) The particular example shown in Figure 1 tends to separate physicists from mathematicians. Try it for yourself before reading the discussion that follows.

---

One of your colleagues is measuring the temperature of a plate of metal placed above an outlet pipe that emits cool air. The result can be well described in Cartesian coordinates by the function

$$T(x,y) = k(x^2 + y^2)$$

where $k$ is a constant. If you were asked to give the following function, what would you write?

$$T(r,\theta) = ?$$

---

*Figure 1: A problem whose answer tends to distinguish mathematicians from physicists.*

The context of the problem encourages you to think in terms of a particular physical system. Physicists tend to think of $T$ as a *physical function* – one that represents the temperature (in whatever units) at a particular point in space (in whatever coordinates). Mathematicians tend to consider $T$ as a *mathematical function* – one that represents a particular functional dependence relating a value to a pair of given numbers.[3]

As a result, physicists tend to answer that $T(r,\theta) = kr^2$ because they interpret $x^2 + y^2$ physically as the square of the distance from the origin. If $r$ and $\theta$ are the polar coordinates corresponding to the rectangular coordinates $x$ and $y$, the physicists' answer yields the same value for the temperature at the same physical point in both representations. In other words, physicists assign meaning to the variables $x$, $y$, $r$, and $\theta$ – the geometry of the physical situation relating the variables to one another.

Mathematicians, on the other hand, may regard $x$, $y$, $r$, and $\theta$ as *dummy* variables denoting two arbitrary independent variables. The variables $(r, \theta)$ or $(x,y)$ don't have any meaning constraining their relationship. Mathematicians focus on the mathematical grammar of the expression rather than any possible physical meaning. The function as defined instructs one to square the two independent variables, add them, and multiply the result by $k$. The result should therefore be $T(r,\theta) = k(r^2 + \theta^2)$.

---

[2] A "shibboleth" is a word or phrase that can be used to distinguish members of a group from non-members.

[3] Mathematicians use an abstract analog of a physical function in classes such as Manifold Theory, but it is only commonly used in math classes at an advanced level.

Typically, a physicist will be upset at the mathematician's result. You might hear, "You can't add $r^2$ and $\theta^2$! They have different units!" The mathematician is likely to be upset at the physicist's result. You might hear, "You can't change the functional dependence without changing the name of the symbol! You have to write something like

$$T(x,y) = S(r,\theta) = kr^2. \tag{1}$$

To which the physicist might respond, "You can't write that the temperature equals the entropy! That will be too confusing." (Physicists often use $S$ to represent entropy.)

Note that we are exaggerating the roles of the mathematician and physicist to illustrate the point that one could have starkly different interpretations of the same expression. Yet, there are times when physicists may pay careful attention to functional form – for example, when considering the transformation from a Lagrangian to a Hamiltonian or between thermodynamic potentials. (Zia et al., 2009) Similarly, a mathematician's back-of-the-envelope work may give preference to quick, intuitive meaning rather than formal syntax. However, in their respective disciplinary cultures, the physicist can use physical justifications for how the mathematical quantities are manipulated or understood, whereas the mathematician's justifications will come from within the mathematical formalisms.

The fact that physicists "load" physical meaning onto symbols in a way that mathematicians usually do not is both powerful and useful. It allows physicists to work with complex mathematical quantities without introducing the fancy math that would be required to handle some issues with mathematical rigor.

### 2.1.2 *Filtering an equation through the physics changes how we interpret it*

A second way that physicists blend physical meaning into math is through "filtering the equation through the physics." The way an equation is used can be strongly affected by the physics it is meant to describe. A nice example is the equation for the photoelectric effect.

This example comes from the modern physics section of the calculus-based physics class for engineers. In the photoelectric effect, electrons bound in a metal by an energy of at least $\Phi$ absorb photons of energy $hf$ where $f$ is the frequency of the light. If the energy (frequency) of the light is high enough, electrons will be knocked out of the metal. As usual, the instructor (EFR) tried to "twist" student expectations a bit by making small but important modifications to standard problems. The problem shown in Figure 3 was a surprise to some of the students but the results were more of a surprise to the instructor.

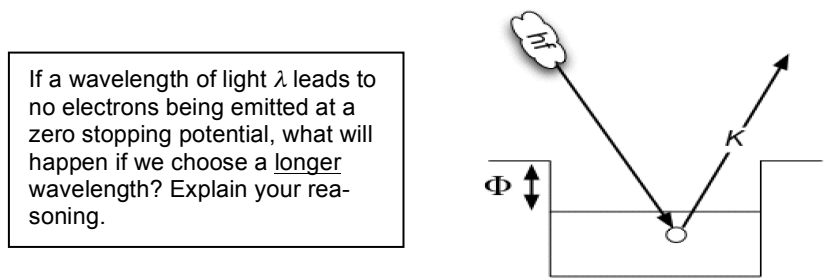

If a wavelength of light $\lambda$ leads to no electrons being emitted at a zero stopping potential, what will happen if we choose a <u>longer</u> wavelength? Explain your reasoning.

*Figure 2: A problem where the equation hides some physics.*

This one is pretty easy if you keep the physics to the fore. A longer wavelength corresponds to a smaller frequency so the photons have *less* energy after the modification. If the photons did not have enough energy to knock out an electron originally, they would certainly not have enough if their energy is reduced even more. Yet nearly a quarter of the class responded that after the modification, there would be electrons knocked out. Their reasoning relied on the Einstein photoelectric effect equation:

$$eV_0 = hf - \Phi \tag{2}$$

where $e$ is the charge on the electron, $V_0$ is the stopping potential, $f$ is the frequency of the photon, and $\Phi$ is the work function of the metal. Their reasoning went as follows: "If $hf$ - $\Phi$ is zero before the change, then it will not be zero after the change. Therefore, since it's not zero there will be electrons knocked out."

This reasoning highlights the fact that Eq. (2) is not actually the correct equation. There is a hidden Heaviside (theta) function that corresponds to the statement: "Do not use this equation unless the right hand side, which corresponds to the maximum kinetic energy of the knocked out electron, is a positive number." A mathematical equation that includes this constraint is

$$eV_0 = (hf - \Phi)\ \theta\,(hf - \Phi). \tag{3}$$

Typically, expert physicists (and introductory physics texts) don't bother to be explicit about this theta function. Instead of encoding the conditions of use in the mathematical formalism, physicists apply their understanding of the physical situation – in this case, checking that the energy of the photon is sufficient to knock out electrons – to evaluate if and how this equation should be used.

### 2.1.3    Interpreting symbols physically yields hidden functions

Because their interest is in mathematical relationships, mathematicians teaching math classes focus on the mathematical grammar of an equation, ignoring possible physical meaning. Novice students often do the same in their physics class, to their confusion. A few years ago, one of us (EFR) gave the example shown in Figure 3 to his second-semester class in algebra-based physics.

A very small charge $q$ is placed at a point $\vec{r}$ somewhere in space. Hidden in the region are a number of electrical charges. The placing of the charge $q$ does not result in any change in the position of the hidden charges. The charge $q$ feels a force, $F$. We conclude that there is an electric field at the point $\vec{r}$ that has the value $E = F/q$.

If the charge $q$ were replaced by a charge $-3q$, then the electric field at the point $\vec{r}$ would be

    a) Equal to $-E$
    b) Equal to $E$
    c) Equal to $-E/3$
    d) Equal to $E/3$
    e) Equal to some other value not given here.
    f) Cannot be determined from the information given.

*Figure 3: A quiz problem that students often misinterpret.*

The instructor had discussed the definition of electric field extensively and he had made it "absolutely clear" that the strength of the electric field was independent of the test charge – making the correct answer (b). Yet on this quiz, more than half of nearly 200 students thought that the answer should be (c). Upon discussing it with the class after the quiz, it became clear that, although many of them could quote the result "the electric field is independent of the test charge," most of the students answering incorrectly had not thought to access that knowledge. They looked at the equation $E = F/q$ and treated the problem as a problem in mathematical grammar. If $E = F/q$, then $F/(-3q) = -E/3$.

In this situation, however, $F$ is not an arbitrary fixed symbol. It represents the Coulomb force that the test charge feels as a result of its interaction with the source charges. It thus is *not* independent of the strength of the test charge. It increases proportionally as the test charge is increased, resulting in the *same* electric field as before.

The instructor *might* have written the force on the test charge as $F_q$ or $F(q)$ in order to remind the students that the force in fact explicitly depended on the test charge (and after this experience he sometimes did so). But the culture of physics expects that each symbol in an equation is to be interpreted in conjunction with its physical meaning. So part of the acculturation of a physics student is learning to interpret the math physically, not to only focus on mathematical manipulations.

## 2.2 Modeling the physical world mathematically

Using mathematical expressions in the disciplines of math or physics is a complex task. In the culture of math, this complexity arises from reasoning and operating in a well-defined and coherent mathematical structure with a particular formal syntax. Yet, the examples in section 2.1 clearly demonstrate that the use of equations in physics goes beyond interpreting and processing the formal mathematical syntax. Instead of relying on explicit Heaviside functions or functional dependences, physicists' use of math is often informed by "checking the physics."

More precisely, in the culture of physics, the use of mathematical expressions is complex, because the ancillary physical meaning of symbols is used to convey information omitted from the mathematical structure of the equation. This is because we have a different purpose for the math: to model real physical systems.

### 2.2.1 Physicists and mathematicians have different goals for the use of math

It's not just that physicists read and use equations differently from the way mathematicians do in math classes. Their goals are different. Physicists don't just want to explore the mathematical formalisms, they want to leverage those formalisms to describe, learn about, and understand physical systems.

In order to explicate the various components of the modeling process for the purpose of thinking about teaching mathematical physics, we use the diagram shown in figure 4. (Redish 2005, Redish & Smith 2008)

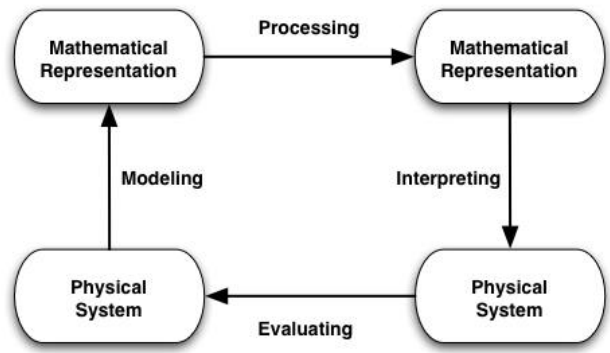

*Figure 4: A model of mathematical modeling*

We begin our description of the modeling process in the lower left by deciding what we are doing. We choose some aspects of a particular physical system we want to describe. We identify measurables –- variables and parameters that we can quantify through some process of measurement (at least in principle, if not always in practice). We then decide what particular mathematical structures are appropriate for describing the features we are interested in. We then *model* the physical system by mapping these measurements into mathematical symbols and expressing the physical-causal relations between the measured quantities in terms of mathematical operations between the symbols.

From the mathematical structures we have chosen, we inherit transformational rules and methodologies for transforming relationships and solving equations. We can then *process* the math to solve problems, leading to answers we were unable to see directly from our physical understanding. This, however, leaves us with only mathematics.

We then have to *interpret* what the result means back in the physical system. Finally, we have to *evaluate* whether the result supports our original choice of model when compared to observations or whether it indicates a need to modify our model.

Each of these four steps – modeling, processing, interpreting, and evaluating – are critical skills in the toolbox of a scientist who uses math to describe the behavior of the physical world. This diagram helps a teacher focus on more than just the mathematical processing that often tends to dominate instruction in physics.

But the use of math in physics is not so simple or sequential as this diagram may tend to indicate. The physics and the math get intimately tangled, as we have seen in the examples above. (And as we'll see in our discussion of students' dynamic cognitive response in section 4.) But the figure serves to emphasize that our traditional way of thinking about using math in physics classes may not give enough emphasis to the critical elements of modeling, interpreting, and evaluating.

Physics instruction tends to provide our students with ready-made models, and we may be exasperated – or even irritated — if students focus on details that we know to be irrelevant. We tend to let them do the mathematical manipulations in the process step, but we rarely ask them to interpret their results and even less frequently ask them to evaluate whether the initial model is adequate.

At the introductory level, our exams often require only one-step recognition, giving "cues" so we don't require our students to recognize deep structures. When they don't succeed on their own with complex problem solving, we tend to pander by only giving simple problems, on which success is not evidence of problem-solving expertise. We often don't recognize what's complex in a problem for a student, and that makes it hard to design appropriate and effective problems.

Our examples show that physicists (as well as other scientists and engineers) often use ancillary physical knowledge — often implicit, tacit, or unstated — when applying mathematics to physical systems. Interestingly enough, a similar idea is valuable to linguists trying to understand how we put meaning to words – semantics. Developments in the linguistics and semantics literature help us to begin build a terminology to be able to better describe the difference between the expectations of the cultures of physicists and mathematicians.

## 3    What do we mean by making meaning? Cognitive Semantics and Mathematics

To understand how we make sense of the language of mathematics in the context of physics, let's consider what is known about how people make sense of language in the context of daily life. Cognitive linguists have made considerable progress on this issue in the past 40 years. Although the research community has not entirely come to an overarching synthesis, they have many ideas that are valuable in helping us make sense of how we make meaning. We offer an exceedingly brief summary of a rich and complex subject, selecting those elements that are particularly relevant. We draw heavily on the work of Lakoff & Johnson (1980), Lakoff & Nunez (2001), Langacker (1987), Fauconnier (1985, 1997), Turner (1991), Fauconnier & Turner (2003), Evans (2006), and the overview of Evans & Green (2006).

From their textbook on modern cognitive semantics, Evans and Green identify core ideas through which meaning is made. We draw on three of these core ideas here (Evans & Green 2006, p. 157):

- *Embodied cognition:* Meaning is grounded in physical experience.
- *Encyclopedic knowledge*: Ancillary knowledge creates meaning.
- *Contextualization*: Meaning is constructed dynamically.

In this section, we will elaborate on how these principles specify how meaning is made in language, and show how those same mechanisms can be used to understand how meaning is made with mathematical expressions.

## 3.1 Embodied cognition: Meaning is grounded in physical experience.

Embodied cognition refers to the interaction of complex cognitive functions with basic physics experience – our sensory perceptions, motor functions, and how they are tied to cultural contexts Infants learn what a shape is by coordinating their vision and touch Toddlers learn the names of these shapes by associating words they hear to objects.

There are abundant examples in everyday language and conceptualization where meaning requires relating to our bodily existence in the three dimensional world that we experience. Lakoff & Johnson (2004) discuss extensive examples of many spatial orientation concepts and metaphors such as "up," "down," "front," and "back" that are tied closely to our spatial experiences:

> ... Thus UP is not understood purely in its own terms but emerges from the collection of constantly performed motor functions having to do with our erect position relative to the gravitational field that we live in. *(Lakoff & Johnson, 2004, p. 56-57).*

This then forms the basis of structuring and understanding more abstract concepts. Metaphorical statements such as: "I'm feeling *up*" or "He is really *low* these days", are conceptualized on the basis of physical orientations.

Lakoff and Johnson (2004) point out how our bodily experiences with physical objects form the basis of conceptualizing emotions such as anger, love, and events such as inflation in terms of discrete entities or physical substances. Such conceptualization forms the basis on which we use and understand statements such as, "Inflation is backing us into a corner," and "You've got too much hostility in you." Many of these are so ingrained and automatic that we do not even realize the metaphorical nature of them in everyday use.

The thesis of embodied cognition states that ultimately our conceptual system is grounded in our interaction with the physical world: How we construe abstract meaning might be constrained by and is often derived from our very concrete experiences in the physical world. Note that embodied cognition is not a reference to the cognitive activity that is obviously involved in performing sensorimotor activities. The idea is that (a) our close sensorimotor interactions with the external world strongly influence the structure and development of higher cognitive facilities, and (b) the cognitive routines involved in performing basic physical actions are involved in higher-order abstract reasoning.

The grounding of conceptualization in physical experience and actions also extends to higher cognitive processes such as mathematical reasoning. Lakoff and Nunez (2001) argue that our understanding of many mathematical concepts relies on everyday ideas such as spatial orientations, groupings, bodily motion, physical containment, and object manipulations (such as rotating and stretching). Thus, the mathematics of set theory can be understood as grounded in our physical experience with containers and collections of objects, and conceptualizing the arithmetic of complex numbers "makes use of the everyday concept of rotation." (p. 29). Many of the sophisticated ideas and formulations in mathematics are intricately entwined with the physicality of our being.

For example, one way that "embodiment" allows mathematics to feel as if it has meaning, even in abstraction, is through *symbolic forms* (Sherin, 1996, 2001, 2006). A symbolic form blends a grammatic signifier – a mathematical *symbol template* – with an abstraction of an understanding of relationships obtained from embodied experience – a *conceptual schema*.

One example of a symbolic form is *parts-of-a-whole*. A symbol template for parts-of-a-whole looks like:

$$\square = \square + \square + \square + \dots \qquad (4)$$

The boxes indicate that any symbol may be in the box – we are only talking about the grammatical structure of the mathematical representation. The conceptual meaning put to it is that something can be considered to be made up of parts. This is something with which we have both direct physical experience and an abstract schema of parts and whole. The symbolic form is considered to be the blend of the symbol template with the conceptual schema: the boxes on the right hand side of the equation each take on the conceptual meaning of "part" and the box on the left hand side takes on the conceptual meaning of "whole," which consists of the sum of all the parts.

Another example of a symbolic form is *base+change*. The symbol template for *base+change* is written as:

$$\square = \square + \Delta \qquad (5)$$

The conceptual structure is that the final amount (box on left hand side) is equal to the starting amount (box on the right hand side) plus how much that starting amount changes by (triangle). Again, this conceptual schema is something with which we have direct physical experience and where we can expect to have abstracted a schema of change.

### 3.2    Encyclopedic knowledge: Ancillary knowledge creates meaning.

The principle of encyclopedic knowledge implies that we understand the meaning of words not in terms of terse definitions provided in a dictionary but in reference to a contextual web of concepts (perhaps represented by other words) that are themselves understood on the basis of still other concepts.

Take the example of the word "hypotenuse," defined as the longest side of a right-angled triangle. To understand the meaning of "hypotenuse," one needs to understand the meaning of a triangle, "sides" of a triangle, right angle, and the idea of "longest". These in turn require conceptualizing a plane, shapes on a plane, lines, angles, length, and so on. Understanding and conceptualizing "hypotenuse" relies on a number of ancillary concepts, which in turn rely on a background of other concepts in an expanding web of encyclopedic knowledge.

This idea of encyclopedic knowledge has been developed under various frameworks such as frame-semantics (Fillmore, 1976), domains (Langacker, 1987), and mental spaces (Turner, 1991). The idea common to these varied perspectives is that we can model knowledge as consisting of a large number of highly interconnected elements. At any given moment, only part of the network is active, depending on the present context and the history of that particular network. The meaning of a word in a particular utterance is then determined by what part of that complex web of knowledge is accessed by that particular utterance of that word. Modern cognitive linguists argue compellingly for these complex links in the structure and processing of meaning:

> …a number of scholars…have presented persuasive arguments for the view that words in human language are never represented independently of context. Instead, these linguists argue that words are always understood with respect to **frames** or **domains** of experience….so that the 'meaning' associated with a particular word (or grammatical construction) cannot be understood independently of the frame with which it is associated. *(Evans & Green 2006, p. 211)*

At first look, the meaning of mathematical equations could seem terse in the dictionary sense. What does it mean to know the meaning of an equation? Consider for example:

$$y = mx + b. \qquad (6)$$

In the strictest sense, it is a statement suggesting a simple calculation, defining the value of the quantity "*y*" in terms of the sum of "*b*" and the product of "*m*" and "*x*". Mathematics sometimes intentionally adopts such a minimalist view. What you know about a mathematical quantity is specified as precisely and as minimally as possible (with axioms), and only that knowledge is to be used in constructing new knowledge.[4]

Though written mathematics can be terse and precise, the culture of math often relies on more than just a "dictionary meaning" in how a symbolic string is used and understood. For most mathematicians (and even high school students) equation (6) carries more meaning than the literal relation between four algebraic and two relational symbols. With a knowledge of labeling conventions, *x* and *y* are interpreted as variables capable of taking on many different values, while *m* and *b* are interpreted as constants. With this addition, the equation takes on the meaning of a relation between the independent variable (*x*) and the dependent variable (*y*). Additionally, the assumed constancy of *m* implies that the equation refers to a straight line. The constants now take on additional mathematical meaning: *m* as the slope of the line and *b* as the intercept on the y-axis, bringing in ideas from graphs. Thus, the meaning of the equation, understood even within the domains of mathematical conventions, straight lines, and graphs are much richer then the strict "definition" expressing the symbol "*y*" in terms of other symbols.

When we use math in physics our rich knowledge about the physical system is additionally brought to bear in interpreting the mathematical meaning as illustrated by the examples in section 2. As another example, think about the equation from physics

$$v = v_0 + at, \tag{7}$$

where *v* is the velocity of an object, $v_0$ is the initial velocity, *a* is the constant acceleration of the object, and *t* is the time. This equation is mathematically identical in grammatical structure to equation (6). But in the context of physics, it is connected to an even richer network of ideas involving motion, speeds, and rates of change. To understand this equation is to understand its relation to all these stores of knowledge. In other words, the meaning of equations in physics is encyclopedic just like the meaning of a word such as "hypotenuse" is encyclopedic. We will see a specific example of how students can perceive this equation in either a mathematical or physical context in section 4.2.1.

### 3.3 Contextualization: Meaning is constructed dynamically.

An implication of the idea that meaning is built by linking to elements of an encyclopedic knowledge system, is that the meaning of entities is not fixed but is dynamically determined based on the specific contextual clues. Semanticists see linguistic units as prompts for the construction of meaning, or, *contextualization*.[5] As described by Evans and Green this is:

> a dynamic process whereby linguistic units serve as prompts for an array of conceptual operations and the recruitment of background knowledge. It follows from this view that meaning is a process rather than a discrete 'thing' that can be 'packaged' by language. Meaning construction draws upon encyclopaedic knowledge…and involves **inferencing strategies** that relate to different aspects of conceptual structure, organization and packaging... [Emphasis in original.] *(Evans & Green, 2006, p. 162)*

If utterances provide access to nodes in the network of knowledge, different parts of which are active in different moments, then the meaning of an utterance depends on the local activity of the network at any given moment. It is in this sense that the meaning is dynamic. The contexts of a particular utterance – local (what is the conversation about, with whom, etc.) and global (personal histories, social and institutional affordances, etc.) – affect the activity of the knowledge network and, in turn, the contextualization at any given moment. Thus, the meaning of an utterance is *not* pre-determined but is constructed in the moment.

---

[4] Although this is the ideal, many mathematicians argue that there are subtle issues that prevent this ideal from ever being realized, even in principle. (Goldstein, 2006; Carroll, 1897).

[5] Evans and Green (2006) refer to this term as *conceptualization*. We have chosen a different term for two reasons: to focus on the fact that this process is the response to context, and to differentiate from the different use of the word *concept* by physics education researchers.

Consider the example of the word "safe" in the following sentences, both referring to a child playing on the beach (Evans & Green, 2006, p.161):

> (1) The child is safe.
> (2) The beach is safe.

The first sentence refers to the child as free from harm. The second sentence, though identical in structure to the first, does not imply that the beach is free from harm. Rather it implies that the beach cannot cause harm. The senses of "safe" in the two sentences (and the properties they assign to child and beach) are slightly different depending on the local context. In a different context, for example, if referring to a beach threatened by property developers, the contextualization and the meaning constructed for the second utterance could be more similar to the first meaning of safe – meaning that the beach is free from being harmed by the developers.

In light of this, we can revisit the example of the imaginary physicist and mathematician arguing about whether $T(r,\theta)$ should be "$r^2 + \theta^2$" or "$r^2$". Although both could likely understand the thinking behind the two responses, each expert makes meaning according to the meanings common to their respective disciplinary cultures. We can now understand the mathematician as attaching meaning to the equation within the domain of simple functions and variables, while the physicist is interpreting the equation within the domain of coordinate systems adding physical meaning to the variables.

Making of meaning with equations shares (at least) three key commonalities with meaning-making in language: an embodied basis, the use of encyclopedic knowledge, and contextual selection of that encyclopedic knowledge for meaning-making. In the next section, we present a theoretical framework used in PER for modeling how students make meaning in physics that aligns with these key features.

# 4    Making meaning in physics: The Resources Framework

The ideas in section 3 tie in nicely with the *resources framework* being developed by the Physics Education Research (PER) community for analyzing student thinking in physics (diSessa, 1988; Hammer, 2000; Hammer, et al., 2004; Redish, 2004; Redish, 2014).[6]  In this framework, an individual's reasoning is modeled as resulting from the activation of a subset of cognitive resources (tightly bound cognitive knowledge structures), activated in response to a perception and interpretation of both external and internal contexts (framing).

This framework, used in PER to understand the dynamics of how students construct meaning in physics, shares many of the key features of frameworks used to understand the dynamics of how individuals construct meaning in language.

---

[6]  As of this writing there are hundreds of papers that use this framework to analyze student thinking about physics. Citations to many of these references can be found online at (Redish & Sayre 2010) or in the KiP Google Community: https://sites.google.com/site/kipcommunity/Home

- *Embodied cognition:* Phenomenological primitives tie basic physics reasoning to embodied experience.
- *Encyclopedic knowledge*: Manifold productive resources are used dynamically
- *Contextualization*: Activation depends on conceptual, epistemological, and affective factors

Here, we provide examples of how the resources framework uses these same core ideas to help us describe the multiple ways in which physics students can make meaning with equations, as well as understand the dynamics of how students can shift from one meaning to another and how conceptual, epistemological, and affective responses can interact in complex ways.

## 4.1 Embodied Cognition:
Phenomenological primitives tie basic physics reasoning to embodied experience.

The resources framework puts its "feet on the ground" in a manner similar to cognitive linguistics – through embodied experience. diSessa (1993) identified basic embodied elements of physics thinking as *phenomenological primitives* ("p-prims"). These are knowledge elements learned, often at a very young age, about how the world works. Two of their core aspects are *obviousness* and *irreducibility* – p-prims are activated easily and directly, and, as far as the user is aware, they have no structure. "That's just the way things are." Two examples (from many cited in diSessa's classic (1993) paper) are "dynamic balancing" (two influences in conflict which happen to balance each other) and "Ohm's p-prim" (an agent or impetus acts through a resistance to produce a result). For example, in thinking about how much a heavy object will move in response to a push, a student may draw on their physical intuition through Ohm's p-prim rather than formal physics principles in their reasoning.

In a similar way, Sherin (2001) found that upper-division physics undergraduates commonly construct novel equations to model physical situations through their intuitive understanding rather than the application of formal physics rules or principles. In describing the forces on an object falling at terminal velocity, a pair of students skipped directly to writing an equation of the form $F_{gravity} = F_{air\ resistance}$. This one-step derivation precludes formal derivation from Newton's laws, where the total force is found to be $F_{net} = F_{gravity} - F_{air\ resistance}$, that is inserted into Newton's second law to give $F_{net} = F_{gravity} - F_{air\ resistance} = ma$, $a$ is taken to be zero for terminal velocity, which gives $F_{gravity} = F_{air\ resistance} = 0$, which means $F_{gravity} = F_{air\ resistance}$. Instead, Sherin models these students' solution as relying on the *balancing* symbolic form to associate the intuitive ideas of two influences in opposition to the symbol template □ = □. This way of leveraging the physical interpretation of the situation to affect how the mathematical equations are generated, used, and interpreted is argued to reflect physics disciplinary expertise and stands in contrast to just formally processing the mathematical syntax.

## 4.2 Encyclopedic Knowledge: Manifold productive resources are used dynamically.

P-prims form a subset of the knowledge that individuals can bring to bear in understanding physical situations. In the resources framework, an individual's knowledge consists of fine-grained knowledge resources. As with how ancillary encyclopedic knowledge is applied to make meaning of language, individuals bring some subset of their resources to make meaning in physics. Different subsets of these resources can be applied to the same situation to form different meanings (just as how "the beach is safe" can be understood to either imply the safety of a beachgoer or the safety of the beach from property development). Because of the manifold possible meanings, learning physics involves refining patterns of activation of and connection to our encyclopedic knowledge base to build a coherent and stable knowledge structure that aligns with the canonical knowledge and reasoning of the discipline of physics.

Although these activations are immensely valuable for living everyday life, sometimes, when mapped into learning situations they can be activated inappropriately or misinterpreted. One example of this is discussed in Redish (2014) (Frank (2009), Frank and Scherr (2012)). Students are given a set of paper tapes with dots on it. (See figure 5.) The dots are made by connecting an object to the paper tape and setting the object in motion. The tape runs through a machine (Pasco Tape Timer) that taps a striking pin at a fixed rate onto the paper tape through carbon paper, making a dot every time the pin taps down as shown at the left in figure 6. If the object is moving quickly, the tape moves a lot between taps and there are large spaces between the dots as shown in the sample at the top right of figure 6. If the object is moving slowly, the tape moves little between taps and the dots are close together as in the bottom right of figure 6.

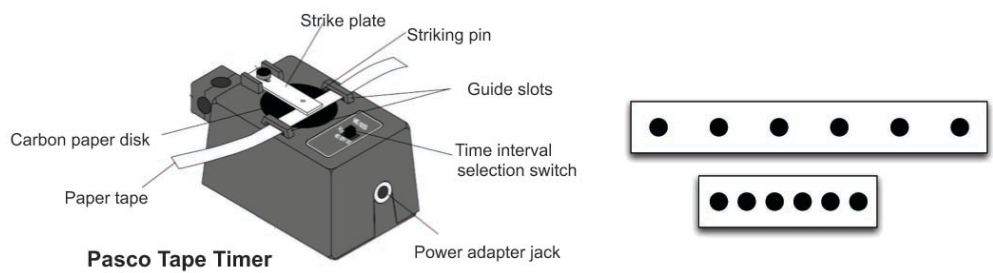

*Figure 5: A device for displaying the speed of motion with spaced dots and two examples.*

The apparatus was shown at the beginning of a recitation section and the mechanism explained by a graduate teaching assistant (TA). Students were working in groups and each group of four students were given four tapes of different lengths containing 6 dots (as shown in figure 6). The first question in their worksheet asked, "Which tape took longer to make?" Since the pin taps at a fixed rate, any tape with 6 dots would take the same time to make. Multiple groups of students were seen to transition quickly from one interpretation to another. In response to the first question in the lesson, asking "Which tape took longer to maker," many students said something like, "Obviously, it takes less time to generate the shorter tape [more closely spaced dots]… because it' shorter." We interpret these students as applying an easily accessible intuitive resource that "more distance implies more time." A short time later, in response to a prompt asking them a more detailed question (to find the velocity), these groups reinterpret the lengths of the strips as proportional to how fast the tape was pulled through the machine, drawing on the intuitive idea that "more distance implies more speed."

There are two important features to note from this common reasoning trajectory. First, in different moments, a group of students can connect different cognitive resources to reach different interpretations about what the different lengths mean. Second, the groups draw on different pieces of their encyclopedic knowledge depending on how they have contexualized the task in response to the different cues in different parts of the lesson to make two different (and mutually exclusive) meanings of the same objects.

A crucial part of the resources framework is the observation that resources are general – neither right nor wrong until the context and use is specified. Although one of the students conclusions about the length of the ticker tapes was incorrect here, their intuitive knowledge resources cannot be defined as correct or incorrect independent of context. In linguistics, it would be wrong to interpret the sentence "Sam broke his arm, so he's in a cast for six weeks" as meaning that Sam was acting in a theater performance given the context. Yet, it would be similarly incorrect to say it is wrong to think "cast" means "a group of actors in the same production." With respect to the ticker tapes, although the intuition "more distance implies more time" is not correct in interpreting these ticker tape strips, it would be correct in other contexts, such as comparing airline flight times between different cities.

In the next section, we provide an example of two ways that physics students can draw on encyclopedic knowledge resources to use and interpret the equation $v = v_0 + at$. We argue that although both meanings are correct ways of interpreting an equation in physics, opportunistically and productively blending physical meaning with the mathematical syntax evidences more expert-like reasoning in the discipline of physics.

### 4.2.1    Case Study #1: Blending physical meaning with mathematical structure

Understanding an equation in physics is not limited to connecting the symbols to physical variables and being able to perform the operations with that equation. An important component is being able to connect the mathematical operations in the equation to their physical meaning and integrating the equation with its implications in the physical world. In this section, we illustrate the differences in ways that meaning could be attached to an equation by analyzing excerpts of clinical interviews with two students, Alex and Pat, in an introductory physics class for engineers (Kuo et al. 2013). The excerpt focuses on the students' understanding of the equation for the velocity of an object moving with a constant acceleration: $v = v_0 + at$, (equation (7)) where $v$ is the velocity of the object at time $t$, $v_0$ is the velocity of the object at $t = 0$, and $a$ is a constant acceleration. Both students can use the equation satisfactorily for solving problems, but the encyclopedic meaning ascribed to the equation by Alex is different than that ascribed by Pat.

When Alex is asked to explain the velocity equation, she focused on the mathematical meaning of the symbols:

> I: Here's an equation; OK, you've probably seen it.
>
> Alex: Yeah.
>
> I: Right. So suppose you had to explain this equation to a friend from class. How would you go about doing that?
>
> Alex: Umm, okay, well, umm, I guess, first of all, well, it's the equation for velocity. Umm, well, I would, I would tell them that it's uh, I mean, it's the integral of acceleration, the derivative of {*furrows brow*} position, right? So, that's how they could figure it out, I don't know. I don't really {*laughs*}, I'm not too sure what else I would say about it. You can find the velocity. Like, I guess it's interesting because you can find the velocity at any time if you have the initial velocity, the acceleration, and time . . .

Alex's response suggests that she thinks about the equation as something that allows one to calculate the velocity of an object at any moment. She does refer to the velocity as the derivative of position and integral of acceleration but her comment does not reflect whether those mathematical operations connecting velocity to position and velocity to acceleration have any deeper physical meaning for her.

Pat also refers to the physical meaning of the symbols in the equation, but her explanation is not limited to their names. Her explanation seems to dip into her physical knowledge about motion, units, and processes of change:

> I: Ok. So here's probably an equation that you have seen before, right? And um, if you were to explain this equation to a friend from class, how would you go about explaining this?
>
> Pat: Well, I think the first thing you'd need to go over would be the definitions of each variable and what each one means, and I guess to get the intuition part, I'm not quite sure if I would start with dimensional analysis or try to explain each term before that. Because I mean if you look at it from the unit side, it's clear that acceleration times time is a velocity, but it might be easier if you think about, you start from an initial velocity and then the acceleration for a certain period of time increases that or decreases that velocity.

Pat also looks at the physical meaning of each symbol in the equation and how they are connected. She brings in knowledge about units and how the dimensions in the equation must be consistent between terms. But she lends a deeper meaning to the equation by bringing in additional knowledge about the physical process of acceleration that changes the initial velocity. Further excerpts of her interview show that, for Pat, the mathematical formulation of equation (7) is stably connected to the conceptual schema of *base+change*, where you start with an initial quantity and then something changes it. The '*at*' term for him reflects the total amount by which the initial value is changed. We model her reasoning as connecting the *base+change* symbolic form to relationships between the physical variables of velocity and acceleration to make meaning of this equation.

In section 2, we argue that the cultural norms of the discipline of physics include opportunistically blending in physical meaning in using and interpreting equations. Because of this, we argue here that Pat's use of *base+change* to blend in an overall conceptual story relating velocity and acceleration onto the velocity equation reflects more-expert-like physics practice.

Furthermore, we see that this interpretation of the equation affects its use in problem solving. Later in the interview, Alex and Pat use the equation to solve a problem about differences in the speeds of two balls thrown from a building at the same time with different initial velocities. Alex uses the equation as a tool to compute final velocities given the initial velocity, time, and acceleration. Pat on the other hand uses the equation much more as an expert physicist might, reaching the answer without needing to plug in numbers and carry out arithmetic calculations, and she exhibits an expert-like understanding of why the result should be what it is based on the meaning that she assigns to the structure of the equation.

### 4.3    Contextualization: Activation depends on conceptual, epistemological, and affective factors.

A human mind contains a vast amount of knowledge about many things but has limited ability to access that knowledge at any given time. (Miller, 1956) As cognitive semanticists point out, context matters significantly in how stimuli are interpreted and this is as true in a physics class as in everyday life. Sometimes, this means that structural features of the problem context cue knowledge related to physics concepts – a problem with a block on a ramp cues the conceptual knowledge of forces and motion rather than conceptual knowledge related to magnetic fields. Yet, context extends beyond the physics content of the particular problems presented in the context of being a student in a physics class interacting with teachers and other students. Students bring a multitude of resources related to "ways-of-knowing," developed through years of experience with schooling and with knowledge building, to physics classes. Since "epistemology" means the science of knowing about knowledge, we refer to these resources as *epistemological resources*.

Elby and Hammer (2001) and Hammer and Elby (2002, 2003) explore what basic ideas people use to decide they know something. Some basic epistemological resources that are relevant for our consideration of mathematics use in science include (Bing & Redish, 2009, 2012):

- *Embodied physical intuition:* Knowledge constructed from experience and perception (e.g., p-prims) is reliable.
- *By authority:* Information from an authoritative source can be trusted.
- *Calculation:* Algorithmic computational steps lead to a trustable result.
- *Physical mapping to math:* A mathematical symbolic representation faithfully characterizes some feature(s) of the physical or geometric systems it is intended to represent.
- *Mathematical consistency:* Mathematical representations have a regularity and reliability that are consistent across different situations.

We can model the encyclopedic knowledge that individuals bring to bear to make contextual meaning as coordinated assemblies of resources – the coordination including conceptual, epistemological, and affective factors among others. As a result of students' experience, both in their everyday lives and in their schooling, we note that often these coordinations of resources may develop into regular, aligned patterns. For example, if mathematical processing is frequently activated with a negative and stressful, *this is hard*, affect, they might tend to activate together.

Although we have thus far emphasized the manifold possible patterns of resource activation, we also believe that sometimes expectations about knowledge and learning can be quite stable and reliably cued – alignment can become strong, at least for a while and in common contexts. In asking how to transform $T(x,y)$ to $T(r,\theta)$, we expect physics and mathematics disciplinary experts to tend to answer according to the rules of their particular disciplinary contexts. In physics, students may spend fifteen minutes or longer, stuck using unproductive knowledge-building methods to solve a problem. (Bing & Redish 2008) Moreover, a student in a physics class may bring to bear ways of making meaning in physics – problem-solving or studying strategies – that consistently fail to be useful, even though there is evidence that they possess the resources for more productive strategies (Hammer, Elby, Scherr, & Redish, 2005). The primacy and stability of particular patterns of aligned activations can inhibit transitions to other potentially more productive ones.

However, it is a mistake to assume that the apparent stability of one line of reasoning implies no alternatives. Just as with language, making meaning in physics is a dynamic process, often responding to shifting social situations and cues. Through the next case study, we make the point that even apparently stable modes of reasoning can be re-evaluated and shifted by an interviewer offering appropriate cues.

### 4.3.1    Case Study #2: How shifts in epistemology supports shifts in the meaning of equations

As an example of the role that epistemology can play in the dynamics of meaning-making in physics, we present the case study of one student's changing interpretations of two "isomorphic" equations (Kuo, 2014).

Devon was interviewed the summer after his first semester physics course. His interview started with the same prompt that Alex and Pat answered: explain the equation $v = v_0 + at$. Devon's explanation, targeted to a middle school student, was similar to Alex's, focusing on the meaning of the variables only as much as required for using the equation as a computational tool.

Yet, we expect that Devon, an undergraduate engineering major, possesses resources for making sense of equations with *base+change* and other symbolic forms. After the velocity prompt, Devon was asked to come up with an isomorphic equation for money:

> You start out with $\$m_0$, and you make $\$r$ per day. How many dollars ($\$m$) would you have at the end of $d$ working days? Could you express the number of dollars ($\$m$) in an equation?

Devon quickly comes up with the correct equation: $m = m_0 + rd$. When asked to explain to a middle school student how he got this equation, he explains through *base+change,* starting with the more concrete example of how much money you would make in one week:

> Devon: So, if you work five days and you get so much money per each day, what do you do to calculate your total earnings for those five days? And I think by twelve they would know, oh just multiply by how much you get per day, OK, and that's going to take care of this, the r times d, and then you know to get the total, they already know what you start off with so they would know to add it to that.

> I: Why add?

> Devon: Because you want to have the total, like for, you start off with a certain amount, and you want to know how much you have after the week, so your initial amount plus how much you earned that week is equal to the total amount of money that you have.

As with Pat's explanation of $v = v_0 + at,$ Devon's explanation of the money equation identifies the quantities of base ("you start off with a certain amount"), change ("how much you earned that week"), and an overall story that relates base and change to final amount ("your initial amount plus how much you earned that week is equal to the total amount of money that you have"). There are many conceptual differences in the problems that contribute to cuing Devon's different interpretations of the two equations. Here, we highlight an epistemological difference: in the interview Devon views understanding equations in physics as memorization and understanding equations in math as understanding why those equations work:

> I: When do you feel really comfortable with an equation, when do you feel that you really understand an equation?

Devon: Well, in physics, I feel comfortable when I memorize the thing, and I know all the units that are attached to it. 'Cuz as I said, I like, I'm a concrete sequential kind of guy in the math, so if I know, if I could see that the units make sense, then I know what I'm doing must be right, I don't, I just don't like thinking of the concepts behind it, I don't like thinking of gravity, you know, Other people think because of gravity, it's going to do such and such, I'm not, you know, I like just focusing on units and just if it makes sense, and I just memorize the equation, I mean, other people can derive the equations by, I don't know, Newton's second law or doing the free body diagrams and they can derive an equation or a certain kinematic-, but I don't do that, I just think of what makes sense, unit wise, I guess.

I: And what about math? When do you feel that you really understand an equation in math?

Devon: Well, in math, well like, there's so many proofs, and it just makes sense in my mind, I don't know, like derivatives and integrals and Jacobian transformations. It just all makes sense to me, because there's a reason it works, and it's just one reason. It's not like in physics really where there's so many different cases like I said before. In math, like if I understand the proofs of why it's that way, and then I'm comfortable using that equation.

Here, two explicit, in-the-moment *epistemological stances* towards equations emerge. In physics, Devon prefers focusing on units and memorizing the equation rather than thinking about the concepts behind the equations. He acknowledges that other people can reason about concepts (like gravity) or derive equations from more basic principles (like Newton's second law), but he explicitly chooses not to.

In contrast, equations in math make sense to Devon. Unlike the multiple different cases in physics, math for Devon is simpler and unified ("There's one reason it works, and it's just one reason."). Perhaps because of the perceived relative simplicity, Devon seeks understanding of math equations whereas the proliferation of equations and concepts in physics make that same kind of understanding seem too difficult to attain.

Later in the interview, Devon moves back to the physics domain to explain an equation for the speed of a car, $s = s_0 + rt$, with *base+change*. This suggests that explaining the money equation not only revealed how Devon makes sense of equations in math problem contexts, but also tapped into a productive resource that he could then, with little effort, apply to make meaning with an equation in a physics context.

Devon's reasoning provides examples of two key features of the resources framework. First, his original failure to use *base+change* does not represent a knowledge deficit. Devon's reasoning with equations is context-dependent and another problem context reveals how he can draw on *base+change* to interpret a similar equation. Second, Devon's epistemological stances towards equations in math and physics align with his reasoning.

### 4.4    Contextualization:
   **How affect shifts along with shifts in conceptual meaning and epistemology.**

Thus far, we have discussed how the dynamics of student reasoning in physics depends on knowledge, both conceptual and epistemological. Yet, as teachers, we know that students' affective states are as important as the knowledge they possess in determining how they make sense of physics problems. For example, as teachers, we know intuitively not to teach new material on the day of an exam because the negative affect that results will impede deep engagement with new topics.

Yet, despite this instructional intuition as well as research showing the correlations between emotions and learning, studies of physics students' conceptual knowledge and epistemologies continue to greatly outnumber the studies that investigate the role of affect and emotions in the dynamics of those students' reasoning. We should not take the detailed specification of conceptual and epistemological resources in the resources framework to imply that other factors, such as affect, play no role in contextualization and meaning-making. Next, we discuss a case study illustrating how affect is coupled with conceptual and epistemological resources in shifts in student reasoning.

### 4.4.1    Case Study #3: Student response to math is dynamic and may be mediated by affect

Gupta and Elby (2011) reported on an extremely interesting interview with a student in an introductory physics class for engineers that illustrates the dynamic response of a student trying to make sense of mathematics in a physics class and displays the interaction of conceptual, epistemological, and affective factors.

The interview occurred during the first semester of physics for engineers. The interviewer was not involved in teaching or evaluating the class. After a discussion of equation (7), a topic that had been discussed in class, the interviewer turned to a topic that had not yet been discussed: pressure in fluids. He presents the equation:

$$p = p_0 + \rho g h \qquad\qquad (8)$$

where $p$ represents the pressure under the surface of a lake, $p_0$ is the pressure at the top of the lake, $\rho$ is the density of water, $g$ is the gravitational field, and $h$ is the depth below the water's surface. He then asks the subject, Jim, whether he has seen the equation before, and Jim replies that he has not. The interviewer then asks whether pressure 7 m beneath the surface of the lake is greater, equal to, or less than the pressure 5 m below the surface.

Jim's first inclination is to use the equation and, after some to-ing and fro-ing he decides that $h$ should be negative under the surface and therefore that the pressure at 7 m below the surface should be less than that at 5 m. He is somewhat uncomfortable with this as it contradicts his intuition and experience. He has made the mistake of taking $h$ as negative (taking positive as upward) but not also taking $g$ as negative (gravity points down).

When asked how a friend from English class might reason about the problem, Jim demonstrates he possesses a good physical intuition about the situation.

> Jim: Like, if they have actually been under water, so the pressure, they might know a little bit about pressure under water…Like a rough estimate. The pressure was higher when I was deeper. [I: Okay] The pressure was lower when I was higher to the surface … They could argue from their personal experience like, one time I was scuba diving and I was like 30 feet below the water and pressure was like, pressure was very high. Like I was just swimming, I was just couple of feet below the water and pressure was not that much.

But he sticks firmly to his trust of the equation. He explains he is in a physics class, states that "this is hard", and expresses the stance that on an exam he would trust his analysis of the equation over his intuition.

> Jim: For an equation to be given to you, it has to be like theory and it has to be fact-bearing. So, fact applies for everything. It is like a law. It applies to every single situation you could be in. But, like, your experience at times or perception is just different—or you don't have the knowledge of that course or anything. So, I will go with the people who have done the law and it has worked time after time after time.

Although Jim can use his embodied experience effectively and correctly, his affect – his lack of confidence about his intuitions and sense that "physics is hard" – leads him to suppress an epistemological resource supporting "knowledge constructed from perception" and to rely instead on knowledge from authority and knowledge constructed from mathematical manipulation. This is shown in figure 6(a) with the lines with arrows indicating activation and the lines with circles on the end indicating inhibition. He appears to have actively rejected his intuition.

I: Do you think the [hydrostatic pressure] equation relates to [the physical experience of pressure]?

Jim: Probably somehow, but not directly. I think there is some way that just completely links the two together, but it's not obvious what that relation is.

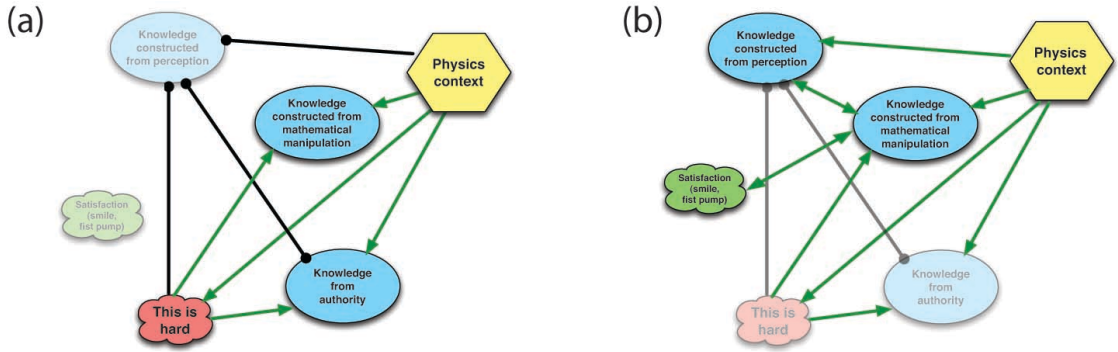

*Figure 6: The interaction of Jim's affect and epistemological resources with the physics context. (a) before reconciliation; (b) after reconciliation. Faded nodes indicate that they are temporarily inactive. (After Gupta and Elby 2011)*

A bit later in the interview, the interviewer hints that perhaps (if he is going to insist that down is the negative direction) that he might consider the sign of *g*.

I: What do you think about *g* in [the hydrostatic pressure] equation? Should that be minus ten or plus ten?

Jim: Oh! minus ten … So, that gives you a positive thing. [I: Okay.] I would say that the negative does not matter anymore. Oooh! I see. The higher you go under water, uh, the lower you go under water the more your pressure is, because the negative and the negative cancel out … So, the more under water you are the higher your pressure is going to be, I think now. I forgot to factor in *g*. That's what I think.

I: Okay. Is that more comfortable or less comfortable?

Jim: That is more comfortable because it actually makes more sense to me now. And now your experience actually does work because from your experience being under water you felt more pressure as opposed to the surface. If I take into consideration both negatives, it is just positive, they just add up.

Jim's epistemological stance has shifted dramatically, along with his affect. Now, his calculation (tied to the epistemological resources "knowledge constructed from mathematical manipulation" and "knowledge from authority") is consistent with his intuition ("knowledge constructed from perception") as shown in figure 6(b).

Overall, the case studies that we have presented in this paper demonstrate that looking at meaning making with mathematics through the lenses of embodied cognition, encyclopedic resources, and contextualization as refined by epistemological analysis gives us insight into ways in which students might not tap into their funds of productive knowledge, how the context can be shifted to reveal that productive knowledge, and how affect plays an interactive role in meaning-making dynamics.

## 5   Implications for instructors

How do we interpret student difficulties with math in physics class? Common diagnoses are that students have not been taught the relevant mathematical tools in their math courses, or they have not learned those mathematical tools well enough to transfer them to their physics courses. The common solutions in either case are for physics instructors to re-teach the needed mathematical procedures and methods in the physics courses or to demand that the students take additional (or more rigorous) math courses.

Our analysis of the roles of physics and math from the point of view of language and meaning making suggests an alternative diagnosis: even if students learned the relevant mathematical tools in their math courses, they still need to learn a component of physics expertise not present in math class – tying those formal mathematical tools to physical meaning..

To succeed in physics, students need not just to be fluent with mathematical processing in the context of physics, but also with mathematical modeling of physical systems, blending physical meaning with mathematical structures, and interpreting and evaluating results. We as physics instructors must explicitly foster these components of expert physics practice to help students succeed in using math in physics.

At a finer-grained level, we are also teaching them how to integrate their everyday embodied and conceptual knowledge with math and how to reliably activate appropriate epistemological resources. Some of these ideas have been discussed in the literature (Tuminaro & Redish 2007, Bing & Redish 2009, 2012, Kuo et al. 2013).

The take-away message is:

> *How mathematical formalism is used in the discipline of mathematics is fundamentally different from how mathematics is used in the discipline of physics – and this difference is often not obvious to students. For many of our students, it is important to explicitly help them learn to blend physical meaning with mathematical formalism.*

The question then is, "How do we help them learn to blend physical meaning with mathematical formalism?" The three case studies we have presented suggest that even when we don't see it, the seeds of this blending are already there. Although we commonly see students failing to use math in physics, case study #1 suggests that some students, like Pat, are already engaging in expert-like use of equations in physics. Case studies #2 and #3 show examples where students who initially reason from the formal procedures of mathematics shift to layering physical meaning onto the math. Importantly, these shifts occurred without direct instruction, suggesting that students possess productive resources for interpreting math physically, which as instructors we need to help them identify and activate reliably.

In other words, to help students integrate physical meaning with the mathematics they are learning and using, we have to use our knowledge of what productive knowledge they possess, what can cue different knowledge elements, what epistemological resources students tend to use, how they feel about physics and math, and how all these elements interact.

While this is easy to say, it can be hard to do. Many physicists have spent years building their fluency and competency with mathematical manipulations until the blending of math with meaning becomes natural and often automatic. For many physicists the easiest way to get through to meaning is to begin with the mathematical relations that come easily to them.

Because of this fluency with blending physical meaning with mathematical manipulations, we have observed physics instructors (including ourselves), teaching at both the introductory and advanced levels, stressing mathematical manipulation as the "go-to" resource for building knowledge. Sample problems done on the board and homework tasks often stress the processing step of the mathematical modeling model described in figure 4, omitting the steps of modeling, evaluating, and interpreting. While for experts this approach may be aligned with "physical mapping to math," for novices who have not developed this fluency with blending the math with physical meaning, the emphasis on mathematical processing can support epistemologies that give primacy to formal laws, equations, and computational reliability to the exclusion of physical meaning.

We argue that instruction not only needs to highlight aspects of how expert physicists layer physical meaning onto mathematics, but that it can tap into productive student resources for doing so Here are two examples.

## 5.1    "Can you do it without the equations?"

In 1992, the senior author of this paper (EFR) decided to switch his research effort to physics education (after 25 years as a theoretical physicist). To learn the ropes, he decided to spend a sabbatical year at the University of Washington, where Lillian McDermott (McDermott 1984) and her Physics Education Group were building the groundbreaking lessons, *Tutorials in Introductory Physics* (McDermott & Shaffer 1992, 2001; Shaffer & McDermott, 1993). As part of his experience in learning how to build lessons that worked, the senior author was being trained as a facilitator for the tutorial lessons. Groups of graduate students, postdocs, and faculty worked through the lesson, discussing how students might respond.

The first week's lesson concerned electric currents and involved the now famous "batteries and bulbs" tasks. The first serious question showed the diagram in figure 7 and asked, "If all the bulbs in the circuit below are identical, rank the brightness of the bulbs."

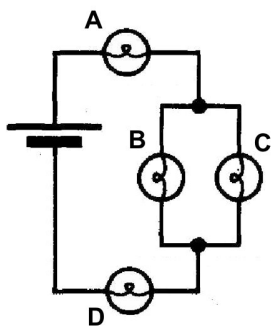

*Figure 7: A batteries and bulbs problem (McDermott & Shaffer 1992).
If the four bulbs are identical, rank the brightness of the bulbs.*

Your experienced theoretical physicist saw no problem with this. "This is a simple Kirchhoff's law problem with two current loops. If I define a current $I_1$ in the left loop and $I_2$ in the right loop, the drops around each loop have to add to the rises, so I get two equations in two unknowns:

$$I_1 R + (I_1 - I_2) R + I_1 R = V$$
$$I_2 R + (I_2 - I_1) R = 0$$

(9)

which are easily solved for the currents. The brightness of each bulb is just the power dissipated, $I^2 R$, so the ranking of the bulbs is like the ranking of the currents. Since my solution gave me $I_1 = 2I_2$, we must have A = D > B = C."

Our theoretical physicist's partner in the task was Richard Steinberg, then a first-year postdoc fresh out of an Applied Physics PhD at Yale. He smiled and asked, "Can you do it without the equations?" Our theoretical physicist frowned, and responded, "Why should I?" Steinberg's reply, "Perhaps your students aren't as fluent with equations as you are", brought a reluctant attempt. The theorist found it surprisingly difficult at first. But after a while, the fog lifted and he realized it was *much* easier without the equations – and that the equations supported and were also interpretable in physical terms. Here's a physical analysis.

Current is conserved so it has to be continuous and divide at junctions (sum of currents in equals sum of currents out). So whatever current goes through the battery has to also go through bulb A. At the junction to the second loop, the current will split. Since both paths have equal resistance, the current will split equally, so the current in B will equal the current in C. At the bottom they come back together and yield the original current. So the current in D is the same as in A – and twice that in B and C. This is the same result as given by the two equations in two unknowns, but it ties directly to the embodied concept of conserved physical flow.

Although experienced physicists will see exactly how this physical analysis connects to the mathematics of the Kirchoff's Law solution, students may not. Recently, one of us (EFR) was serving in the help center when a student asked for help with a circuit problem of the same level of complexity as the one shown above. He had struggled for an hour trying to set up and solve the Kirchhoff's law loop equations as he had been taught in class. When he was introduced to the conceptual approach to using the basic principles he lit up and was able to solve the problem quickly and easily, saying, "Why weren't we shown this way to do it?" He would still need to bring his conceptual understanding into line with the mathematical reasoning needed to set up more complex problems, but the conceptual base made sense to him as a starting point in a way that the algorithmic math did not.

Instructionally, for many students the first step needs to be highlighting the physical meaning, as Steinberg did by asking "can you do it without the equations?" However, the instruction is incomplete until these physical interpretations are tied back to the formal mathematical laws. For example, the conceptual idea that currents are continuous and that the voltage drop across each resistor should sum to the voltage supplied by the battery is represented mathematically through the procedure of writing the voltage changes around a closed loop and by interpreting the final equation with the *parts-of-a-whole* symbolic form.

The benefits of this physical meaning here are two-fold: (1) the procedure of Kirchoff's Law analysis is no longer a brittle, rote procedure, but is backed up by a physical understanding, and (2) it may yield conceptual shortcuts in complicated circuits where setting up the equations for multiple loops may be more challenging.

## 5.2    Tapping into physical intuition for understanding mathematics

The senior author has also had multiple opportunities to observe the instruction of a number of physics faculty. A particularly interesting example occurred in a class of mostly biology majors in which students were learning to read potential energy graphs and relate them to forces. The students were shown figure 8(a) and asked the question: "If two atoms have a total energy $E$ shown by the red line, and the potential energy interaction is as shown by the curve in the figure, when they are at a separation indicated by C, is the force between them attractive or repulsive?"

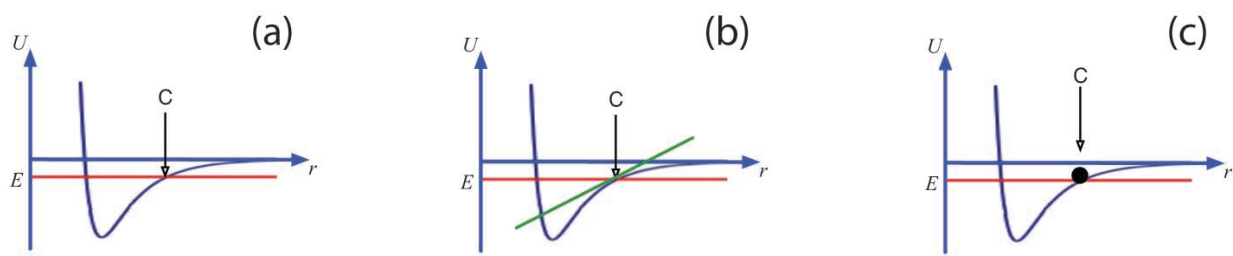

*Figure 8: Three figures illustrating different epistemological approaches to an explanation.*
*(a) The figure shown in the problem; (b) an explanation based on a formula;*
*(c) an explanation based on a physical analogy.*

The students were still at an early stage in learning to read potential energy diagrams and struggled with the question. In two different classes two different instructors essentially responded with the argument, "Remember the equation shown on the board [Eq. (10)]. Since the derivative of the PE curve at C is positive [the slope of the green tangent line in figure 8(b)], the force is negative, which means it's attractive." (Both instructors wrote equation (10) on the board, but neither actually drew figure 8(b).)

$$F = -\frac{dU}{dr} \tag{10}$$

Again, for students who haven't developed facility in reading physical meaning from mathematical expressions, they may make sense of this lesson with the epistemological resources, *by trusted authority* (the remembered equation) and *calculation*.

A more effective approach for this population might be to begin with direct physical experience or an embodied analogy, and implicitly supporting an epistemologies valuing physical intuition. Start with treating a potential energy curve as a track or hill and using the analogy of gravitational PE, then place a ball on the hill as shown in figure 9(c). Which way will it roll? In conversations with a number of these students, none had any difficulty in coming up with the correct answer. Then, inferring that the force was to the left, they were able to see that the force to the left implied a positive slope for the PE via equation (10) and were then easily able to explain how that could be seen from the graph and the equation.

### 5.3    Teaching physics standing on your head.

In the two examples discussed in sections 5.1 and 5.2, one could model these *physicist's* active epistemologies in the two competing instructional sequences as connected chains of epistemological resources, as displayed in figure 9. The one at the left starts from valuing formalized equations and principles. The one at the right gives primacy to physical intuition and embodied experience. In many instructional situations (especially with novice or non-physics major students) the sequence on the right can be more effective for supporting the blending of physical meaning and mathematics, despite the comfort that an experienced physicist might have with the one on the left. There are two reasons for this.

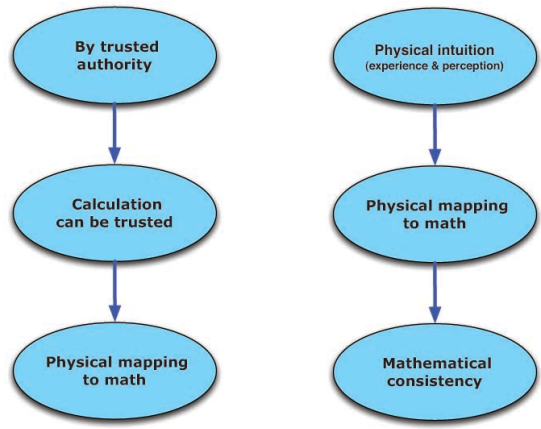

*Figure 9: Two plausible models of epistemological resources activated for the examples in 5.1 and 5.2. The one on the left models what resources physicists may activate when starting from formal equations. The one on the right models a chain may be more useful in helping introductory students learn the value of math in science.*

The first is that we want to provide students with experience with the physics disciplinary practice of physically interpreting mathematics. Starting from the formal laws of physics, while efficient, may not make the underlying physical meaning in those expressions visible to novices. From the two examples above, starting from the physical meaning and then explicitly mapping this meaning to the mathematics can help make this connection explicit for physics students (as well as instructors) for particular topics and help students see how to make this connection more generally.

The second is that, in physics, we want students to learn to take an epistemological stance that coherence between physical meaning and mathematical formalism is valuable and productive. Although in particular moments experts may appeal to formalized physics laws and equations, we believe that that same experts, in other moments, regularly blend physical meaning with the mathematics. However, novice stu-dents, even those capable in math, have to learn to the language of associating physical meaning with math. Because the connection between the physical meaning and the mathematics is more opaque (especially for novices) in the first sequence, students may see it as an example where physics class values knowledge from authority over physical intuition. This has the instructional consequence of encouraging a novice epistemological stance (memorizing equations without understanding) even while facilitating and quickly generating a "correct" answer.

In both of our examples in this section, experienced physicists' epistemologies could be plausibly modeled through the sequence on the left, and, for them, we expect that this sequence is a stable and reliable one. This is likely because this sequence can be very productive in a physics researcher's career. Yet, the stability of this instructional sequence may also work to inhibit the activation of other sequences in instruction that may better support the student epistemologies we believe are productive for making sense of math in physics.

It might well be preferable to "teach physics standing on your head" by beginning with the physical meaning and creating a chain of association to the math, both strengthening the students' skills of "seeing physical meaning" in equations and helping them develop the epistemological stance that equations in physics should be interpreted physically.

We suggest that an analysis of instructional success and activity building using this sort of analysis might prove productive in research and in the building of instructional materials. (See, for example, Elby 2001, Gupta & Elby 2011, Bing & Redish 2012, and Kuo et al. 2013.)

# 6  Conclusion

In a standard interpretation, lack of success with math in physics is attributed to the failure of transfer of mathematical skills from math class to physics class. We advocate for an alternative diagnosis: learning math in math class and math in physics class should be treated as learning two related but distinct languages: although there is significant overlap, there are also important differences, and expertise in one does not guarantee expertise in another. We must recognize that we are asking students to become bilingual and provide instruction and practice in blending in physical interpretation on the mathematical formalism.

# Acknowledgments

We are grateful for discussions with Andrew Elby and Ayush Gupta on the topic of this paper. We would like to thank the members of the University of Maryland Physics and Biology Education Research Groups for many relevant conversations.